\newcommand{\vectr}{\bm{r}}

\newcommand{\phistar}{\phi^*}
\newcommand{\phirtau}{{\phi}(\vectr, \tau)}
\newcommand{\phistarrtaup}{{\phi}^*(\vectr, \tau+)}

\newcommand{\omegabar}[0]{\bar{\omega}}
\newcommand{\omegabardot}[0]{\dot{\bar{\omega}}}

\newcommand{\Ekin}{E_{\mathrm{kin}}}
\newcommand{\Etot}{E_{\mathrm{tot}}}
\newcommand{\Erel}{E_{\mathrm{rel}}}

\documentclass[%
reprint,
superscriptaddress,
longbibliography,
 amsmath,amssymb,
 aps,
]{revtex4-2}

\usepackage{graphicx}
\usepackage{dcolumn}
\usepackage{bm}
\usepackage{amsmath}
\usepackage{physics}
 \usepackage{hyperref}
 \usepackage{xspace}
 \hypersetup{ 
    colorlinks=true,
    linkcolor=blue,     
    urlcolor=blue,
    citecolor=blue,
    }
\newcommand{\cyclestep}[1]{${#1}$\xspace}
\newcommand{\sta}{\cyclestep{A}}
\newcommand{\stb}{\cyclestep{B}}
\newcommand{\stc}{\cyclestep{C}}
\newcommand{\std}{\cyclestep{D}}

\begin{document}

\preprint{APS/123-QED}

\title{Thermodynamic engine with a quantum degenerate working fluid}

\author{Ethan Q. Simmons}
\author{Roshan Sajjad}
\affiliation{Department of Physics, University of California, Santa Barbara, California 93106, USA}

\author{Kimberlee Keithley}
\affiliation{Department of Chemical Engineering, University of California, Santa Barbara, California 93106, USA}

\author{Hector Mas}
\author{Jeremy L. Tanlimco}
\author{Eber Nolasco-Martinez}
\author{Yifei Bai}
\affiliation{Department of Physics, University of California, Santa Barbara, California 93106, USA}

\author{Glenn H. Fredrickson}
\affiliation{Department of Chemical Engineering, University of California, Santa Barbara, California 93106, USA}

\author{David M. Weld}
\affiliation{Department of Physics, University of California, Santa Barbara, California 93106, USA}


\begin{abstract}

Can quantum mechanical thermodynamic engines outperform their classical counterparts? To address one aspect of this question, we experimentally realize and characterize an isentropic thermodynamic engine that uses a Bose-condensed working fluid. In this  engine, an interacting quantum degenerate gas of bosonic lithium is subjected to trap compression and relaxation strokes interleaved with strokes strengthening and weakening interparticle interactions. We observe a significant enhancement in efficiency and power when using a Bose-condensed working fluid, compared to the case of a non-degenerate gas. We demonstrate reversibility, and measure power and efficiency as a function of engine parameters including compression ratio and cycle time. Results agree quantitatively with exact interacting finite temperature field-theoretic simulations. 

\end{abstract}

\maketitle

Classical thermodynamic engines have been critical to human technology since the industrial revolution. In the past decade, the capabilities of quantum thermodynamic engines have been explored theoretically ~\cite{quan_quantum_2007,zheng_work_2014,zheng_quantum_2015,kosloff_quantum_2017,hamedani_raja_finite-time_2021,beau_scaling-up_2016,li_efficient_2018,chen_interaction-driven_2019,yunger_halpern_quantum_2019,barontini_ultra-cold_2019,brandner_nonequilibrium_2020,keller_feshbach_2020,gluza_quantum_2021,boubakour_interaction-enhanced_2023, eglinton_thermodynamic_2022, myers_boosting_2022, rosnagel_nanoscale_2014, zhang_quantum_2014, zheng_work_2014, park_quantum_2019, wu_quantum_2006}, and recent years have seen experimental demonstrations of both quantum and nanoscopic classical engines using single ions~\cite{single_atom_2016,bu_enhancement_2023}, nuclear spins~\cite{peterson_experimental_2019}, cold atoms~\cite{brantut_thermoelectric_2013,QHE_EIT_2017,nettersheim_power_2022}, nitrogen-vacancy centers~\cite{klatzow_experimental_2019}, and quantum gases ~\cite{bouton_quantum_2021,koch_making_2022}. A natural question is whether quantum phenomena can enhance the performance of a thermodynamic engine~\cite{park_efficiency_bound_2019, watanabe_quantum_2020,jaramillo_quantum_2016}. Perhaps the simplest experimental approach to this question --- the direct comparison of an engine using a classical working fluid to an equivalent one using a quantum degenerate working fluid --- has remained unexplored.

\begin{figure}[th!]
\centering
\renewcommand{\figurename}{Fig.}
\includegraphics[scale=.88]{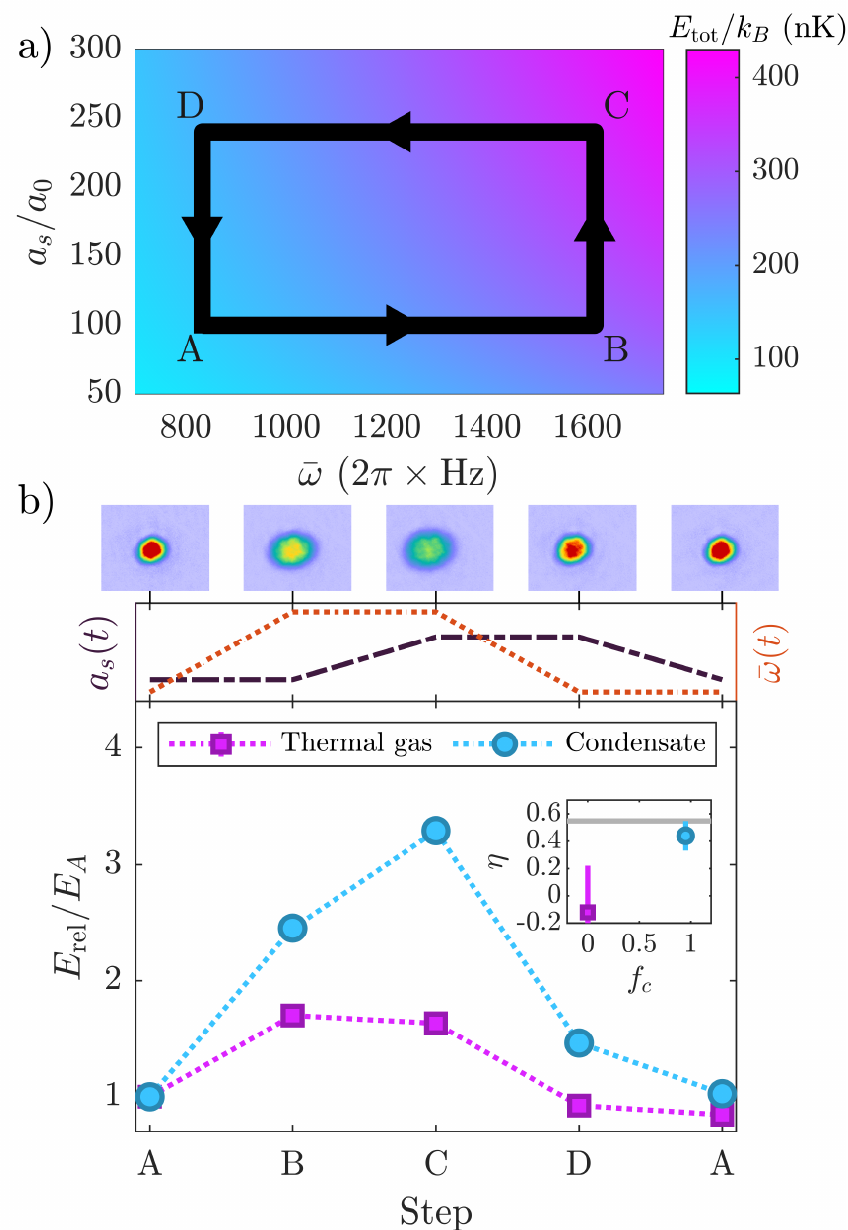}
\caption{Thermodynamic engine with a quantum degenerate working fluid. \textbf{(a)} Engine cycle in $a_s-\omegabar$ space. Color shows total energy per particle. \textbf{(b)} Top: BEC images after 12 ms of expansion at each step. Middle: Evolution of trap frequency (dotted) and scattering length (dot-dashed). Bottom: measured release energies for quantum degenerate (circles) and thermal (squares) working fluids during one engine cycle, normalized by the step \sta~value. Dotted lines connect data points. Inset shows efficiency for each condensate fraction $f_c$; line indicates theoretical maximum efficiency in the Thomas-Fermi regime. Error bars show standard error in all figures.}
\label{fig1}
\end{figure}

In this work, we experimentally realize and characterize an isentropic thermodynamic engine with a quantum degenerate working fluid. The engine cycle interleaves compression and decompression of an optical trap with Feshbach enhancement and suppression of interparticle scattering to pump energy from a magnetic field to an optical field via a trapped ensemble of ultracold neutral lithium.  We observe that quantum degeneracy significantly enhances the output power.  We measure the dependence of efficiency and power on cycle time, and investigate the effects on engine performance of compression ratio and interaction strength ratio. Results agree quantitatively with both approximation-free finite temperature interacting numerical simulations and mean-field analytics.

The experiments begin by preparing a Bose-Einstein condensate (BEC) of 300,000 to 1 million $^7$Li atoms in a far-detuned crossed optical dipole trap with a mean trap frequency $~\omegabar=2\pi \times 133$ Hz, at a temperature of 170 nK, corresponding to a condensate fraction of 0.95. After evaporative cooling to degeneracy,  the $s$-wave interparticle scattering length is Feshbach-tuned to 100$a_0$, where $a_0$ is the Bohr radius.  This sets the initial condition (labeled $A$ in figures).  Interleaved variation of the trap intensity and Feshbach field then execute the thermodynamic cycle illustrated in Fig.~\ref{fig1}a. Between steps $A$ and $B$ (stroke $AB$), the trap power is increased with a functional form such that  $\bar{\omega}$ increases from $\omegabar_A$ to $\omegabar_B$ at a constant rate. This is the compression stroke of the engine. In stroke $BC$, the trap frequency is held constant as the interaction strength is ramped from $a_s^B=100a_0$ to a larger value $a_s^C$ at a constant rate. 
Subsequently, the trap frequency and then the interactions are ramped linearly back to their initial values. Such a cycle pumps energy between the magnetic and optical control fields, because the work performed by the strongly interacting gas during decompression is not equivalent to the work done to compress the more weakly interacting gas. Performing the strokes of the cycle in the order shown in Fig.~\ref{fig1}a results in a net transfer of energy from magnetic to optical fields. Appendix B details an intuitively useful analogy between this isentropic cycle and the Otto heat-engine cycle.

The second-quantized Hamiltonian describing the working fluid includes kinetic, interaction, and potential terms: 
\begin{gather}
    \hat{H}=\hat{H}_{\mathrm{kin}}+\hat{H}_{\mathrm{int}}+\hat{H}_{\mathrm{pot}},\\
    \hat{H}_{\mathrm{kin}}=\int\left(\frac{\hbar^2}{2 m} \nabla \hat{\Psi}^{\dagger} \nabla \hat{\Psi}\right) \mathrm{d}^3 r,\nonumber\\
    \hat{H}_{\mathrm{int}}=\frac{g(t)}{2} \int \hat{\Psi}^{\dagger} \hat{\Psi}^{\dagger} \hat{\Psi} \hat{\Psi} \:  \mathrm{d}^3r,\nonumber\\
    \hat{H}_{\mathrm{pot}}=\sum_{\boldsymbol{\mathrm{k}}}\hbar\omega_{\boldsymbol{\mathrm{k}}}(t)\left(\hat{\Psi}_{\boldsymbol{\mathrm{k}}}^\dagger\hat{\Psi}_{\boldsymbol{\mathrm{k}}}^{{\color{white}\dagger}}+\frac{1}{2}\right), \nonumber
    \label{hamiltonian}
\end{gather}
where $g(t)=4\pi \hbar^2 m^{-1} a_s(t)$ is the interaction coupling constant, $a_s(t)$ is the scattering length, $m$ is the mass, and $\omega_{\boldsymbol{\mathrm{k}}}(t)$ is the trap frequency of mode \boldsymbol{\mathrm{k}} at time $t$. To measure release energy (defined below) at each step, we first abruptly switch off the trap, quenching to zero the last term of the Hamiltonian  $\hat{H}_{\mathrm{pot}}$. Following 12 ms of free expansion we measure the column-integrated density distribution by absorption imaging and reconstruct the 3D distribution via Abel inversion~\cite{hu_bose_2016}. After expansion, not only is the initial momentum distribution converted to a position distribution, but also essentially all the initial interaction energy is converted to kinetic energy~\cite{dalfovo_RMP_1999}, so the distribution provides a measure of the condensate's release energy $E_{\mathrm{rel}}=E_{\mathrm{kin}}+E_{\mathrm{int}}$. We report release energies per atom, while plotted powers represent the total engine power.

Engine performance can be characterized by efficiency and power. We define work done on the condensate as positive. As in refs.~\cite{li_efficient_2018,keller_feshbach_2020}, we define the efficiency 
\begin{equation} \label{eqn:effdefinition}
    \eta=-\frac{W_{AB}^\mathrm{las}+W_{CD}^\mathrm{las}}{W_{BC}^\mathrm{mag}}, 
\end{equation}
and the power
\begin{equation} \label{eqn:powerdefinition}
    P=-\frac{W_{AB}^\mathrm{las}+W_{CD}^\mathrm{las}}{T_{\mathrm{cycle}}}.
\end{equation}
Here $W_{ij}^k$ is the work done on the BEC by the field $k$ (laser or magnetic) in stroke $ij$ of the  cycle and $T_{\mathrm{cycle}}$ is the total cycle time. 

While we measure the release energy rather than the total energy, thse quantities can be simply related via the Gross-Pitaevskii description of an interacting gas. In the Thomas-Fermi regime, the total energy is given by~\cite{pethick_smith_2008}
\begin{align} \label{eqn:TF_energy_balance}
E_{\mathrm{tot}} &= E_{\mathrm{kin}} + E_{\mathrm{pot}}+E_{\mathrm{int}} = \frac{5}{7}\mu,
\end{align}
where $\mu$ is the chemical potential, $E_{\mathrm{pot}}=(2/7)\mu$, $E_{\mathrm{int}}=(3/7)\mu$ and $E_{\mathrm{kin}}\approx 0$. The ratio between the total energy and the measured energy is then
\begin{align}\label{eqn:power_factor}
    \frac{E_{\mathrm{tot}}}{E_{\mathrm{rel}}}  &\approx \frac{E_{\mathrm{pot}}+E_{\mathrm{int}}}{E_{\mathrm{int}}}
    =2.5.
\end{align}
Therefore, a power measurement based on release energy will be reduced from the true power by a factor of 2.5, while measured efficiency will give the true value. As shown later, we have verified the validity of these assumptions using approximation-free numerics.

Fig.~\ref{fig1}b demonstrates the stark contrast between the behavior of degenerate and non-degenerate gases subjected to similar thermodynamic cycles. 
The thermal gas is prepared via inhibited evaporation at a small scattering length of 57$a_0$, resulting in a density of $\small{\sim} 6\times10^{11}\,\mathrm{atoms/cm^3}$ at a temperature of 890 nK. The density of the condensate is $\small{\sim} 2\times 10^{13}\,\mathrm{atoms/cm^3}$, about 33 times larger, at a temperature of 170 nK. Much of this enhancement in density is a direct result of bosonic quantum statistics. While the thermal gas and condensate are prepared at different trap frequencies, the compression ratio $\nu=\bar{\omega}_B/\bar{\omega}_A \approx 2$ is the same for both. As interaction strength increases, the low density of the thermal gas results in a negligible change in release energy, while the Bose-enhanced density of the quantum degenerate sample results in a significant change. The measured efficiency of the engine with a thermal working fluid is consistent with zero, while the measured efficiency of the engine with a quantum degenerate working fluid is $0.45\pm0.1$, near the maximum theoretical value of 0.55 for this compression ratio.  

\begin{figure}[t]
\centering
\renewcommand{\figurename}{Fig.}
\includegraphics[scale=1]{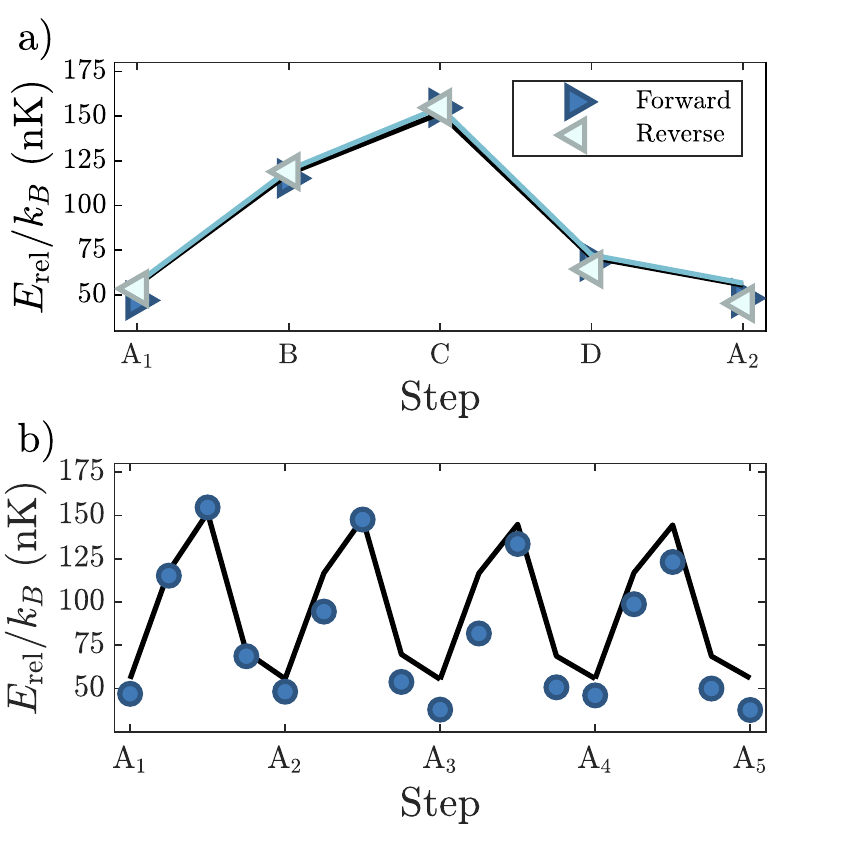}
\caption{Engine reversibility and repeatability. \textbf{(a):} Comparison of the cycle performed in the ``forward'' (\sta-\stb-\stc-\std-\sta) and ``reverse'' (\sta-\std-\stc-\stb-\sta) directions, indicated by right- and left-pointing markers respectively. Light blue line shows results of analytic calculations (see Eq.~\ref{eqn:erel_analytical}); black line shows results of isentropic fully-interacting numerical simulations in both panels. \textbf{(b):} Measured release energy evolution during four repeated engine cycles. Simulation particle number is set to the mean particle number across each four-step cycle. Error bars are smaller than symbol size.
}
\label{fig2}
\end{figure}

Reversibility can be tested by comparing the results of forward (\sta-\stb-\stc-\std-\sta) and reverse (\sta-\std-\stc-\stb-\sta) cycles. Fig.~\ref{fig2}a shows the experimental results of such a  comparison, demonstrating a high degree of reversibility and confirming that the reverse cycle results in a net transfer of energy from optical to magnetic fields, opposite to the forward cycle. Fig.~\ref{fig2}b shows that the same cycle can be performed many times. The repeated return of the condensate to its initial release energy indicates that it can mediate energy transfer between magnetic and optical fields without significant net absorption of energy. 

\begin{figure}[t]
\centering
\renewcommand{\figurename}{Fig.}
\includegraphics[scale=1]{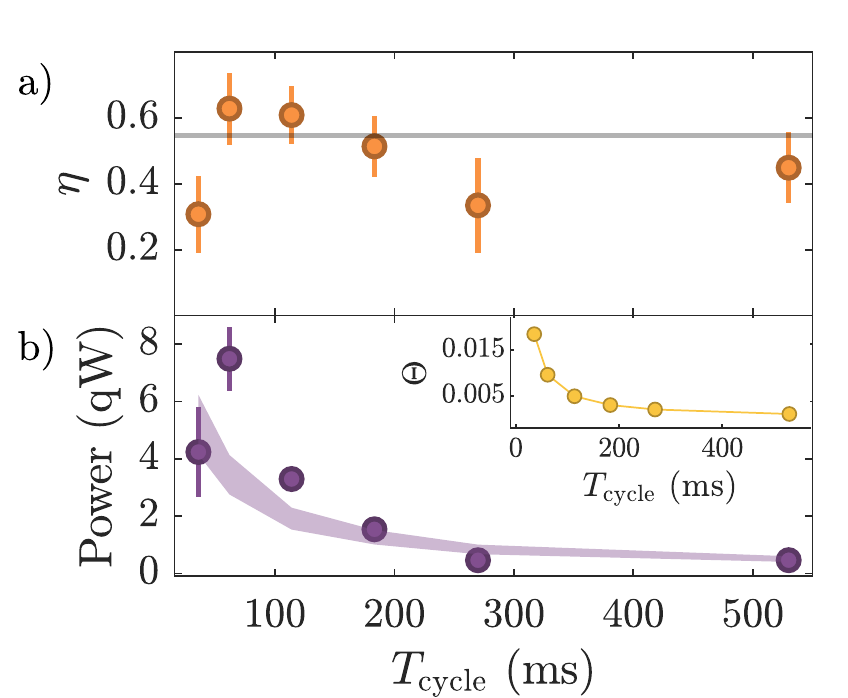}
\caption{Efficiency and power vs. cycle time. \textbf{(a):} Measured energy transfer efficiency $\eta$ versus cycle time.  Line shows theoretical efficiency from Eq.~\ref{eqn:erel_analytical}. \textbf{(b):} Measured engine power, quoted in quectoWatts ($10^{-30}$~Watts), versus cycle time. Shaded region shows the theoretical prediction of Eq.~\ref{eqn:erel_analytical} for the measured range of atom numbers. The power shown here is taken from release energy measurements; as discussed in the main text, the total  power is a factor of 2.5 higher. Inset shows adiabaticity parameter $\Theta$ versus cycle time.}
\label{fig3}
\end{figure}

To estimate the degree of adiabaticity, one can apply the Landau-Zener formalism~\cite{zener_avoided_1932} to approximate the probability of low-lying collective excitations~\cite{stringari_excitations_1996}. Considering only the ground state and lowest-lying collective excitation, the probability of diabatic passage between them is $P_D = \exp(-1/\Theta)$,  where the adiabaticity parameter $\Theta=\dot{\omega}_E/(2\pi\omega_E^2)$ depends on the energy gap $\hbar \omega_E$ to the nearest excited level. Taking our trap to be approximately axially symmetric, and using the results of~\cite{stringari_excitations_1996} with a known ramp speed $\omegabardot=2\pi \times 1$ Hz/ms, we estimate a maximum adiabaticity parameter of $\Theta \simeq0.001$ for the cycles shown in Figs.~\ref{fig1} and \ref{fig2}, with cycle times of 530 ms. 

Varying the engine cycle time affects both efficiency and power. Fig.~\ref{fig3}a compares measured efficiency to the ideal Thomas-Fermi efficiency, which is independent of cycle time.  Measurements for a range of cycle times cluster near this ideal. However, at long cycle times three-body loss, one-body loss, and heating can degrade efficiency, while at the shortest cycle times a combination of technical limitations (for example inductive limits on magnet current ramp rate) and decreasing adiabaticity affect engine performance.

Measuring engine power, we observe the expected inverse dependence over a range of cycle times, as shown in Fig.~\ref{fig3}b.
Power increases for faster cycles, deviating somewhat from the adiabatic prediction of Eq.~\ref{eqn:erel_analytical} as the cycle time is reduced. The breakdown of engine performance at very short cycle times is also visible. These results indicate an optimal range of working speeds; as with any engine, there is a balance to be struck between power and efficiency~\cite{brandner_coherence-enhanced_2015}. Related theoretical work has explored the possibility of bypassing this trade-off using shortcuts to adiabaticity~\cite{li_efficient_2018,beau_scaling-up_2016,schaff_shortcut_2011}.

To investigate the validity of our theoretical approximations, we compare experimentally measured release energies to the results of fit-parameter free, finite temperature equilibrium simulations reproducing the experimental particle number, scattering length, and confinement. The engine's demonstrated reversibility and adiabaticity justify the use of multiple equilibrium simulations and an assumption of isentropic evolution. At steps \sta through \std, we model the system as an interacting confined Bose gas using a path integral over complex-conjugate coherent states fields $\phi$ and $\phi^*$ with the action given in continuous imaginary time notation as~\cite{Orland_book}
\begin{multline}\label{eqn::action}
 S = \int_0^{\beta} d {\tau} \int d^d r \left\{  \phistarrtaup \left[\partial_{ \tau} - \hbar^2/(2m)  \laplacian  \right. \right. \\ 
    \left. \left. + {U}_\mathrm{ext}(\vectr) -  \mu  \right] \phirtau + \frac{ g}{2} \left[ \phirtau \phistarrtaup \right]^2 \right\},
\end{multline}
 where the notation $\tau+$ indicates the field should be evaluated at an advanced position on the $\tau$ contour, with $\tau \in [0, \beta]$ and $\beta = 1/k_BT$. $U_\mathrm{ext}(\vectr) = \frac{1}{2} m (\omega_x^2 x^2 + \omega_y^2 y^2 + \omega_z^2 z^2)$ is the confinement potential, with $\omega_i$ the angular trap frequency in the $i^\mathrm{th}$ direction.  Interactions are modeled as pairwise contact repulsions. The chemical potential $\mu$ is constrained such that total particle number $N$ is constant in each simulation~\cite{delaney_numerical_2020}. 
 We sample configurations of this field theory using the complex Langevin (CL) technique, a stochastic method of evaluating integrals that is robust for actions with a sign problem~\cite{parisi, klauder} such as that in Eq.~\ref{eqn::action}. Observables are calculated by time averaging field operators, obtained from thermodynamic derivatives of the partition function. This method does not require simplifying approximations, even at finite temperature, so it fully accounts for quantum and thermal fluctuations. The use of fields rather than particle coordinates allows full-scale replication of the experimental system on readily-available GPU hardware. Further details of the numerical methods appear in Appendix A.

\begin{figure}[t]
\centering
\renewcommand{\figurename}{Fig.}
\includegraphics[scale=1]{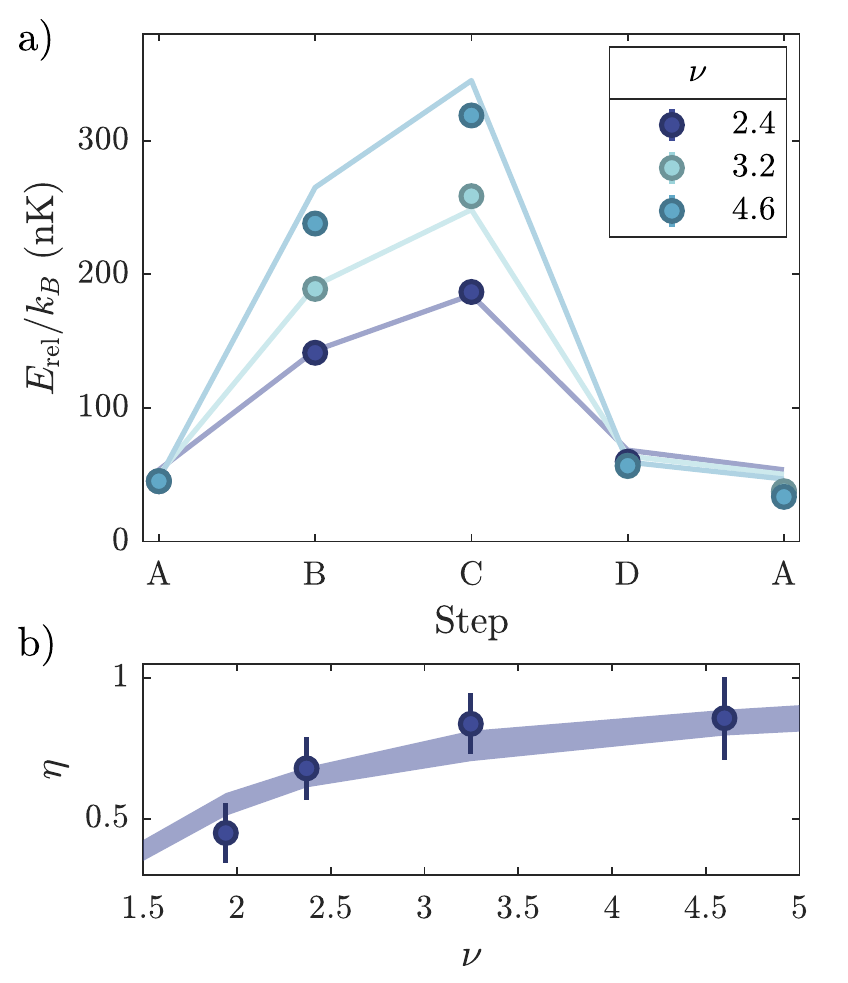}
\caption{Varying compression ratio. \textbf{(a):} Measured release energy evolution over one engine cycle for varying $\nu=\bar{\omega}_B/\bar{\omega}_A$ at a fixed interaction ratio $\kappa=a_s^C/a_s^A=2.4$. Lines show analytical prediction of Eq.~\ref{eqn:erel_analytical}. \textbf{(b):} Efficiency $\eta$ as a function of compression ratio. Shaded region shows theoretical prediction of Eq.~\ref{eqn:effcomp} for the measured range of atom numbers.}
\label{fig4}
\end{figure}

\begin{figure}[t]
\centering
\renewcommand{\figurename}{Fig.}
\includegraphics[scale=1]{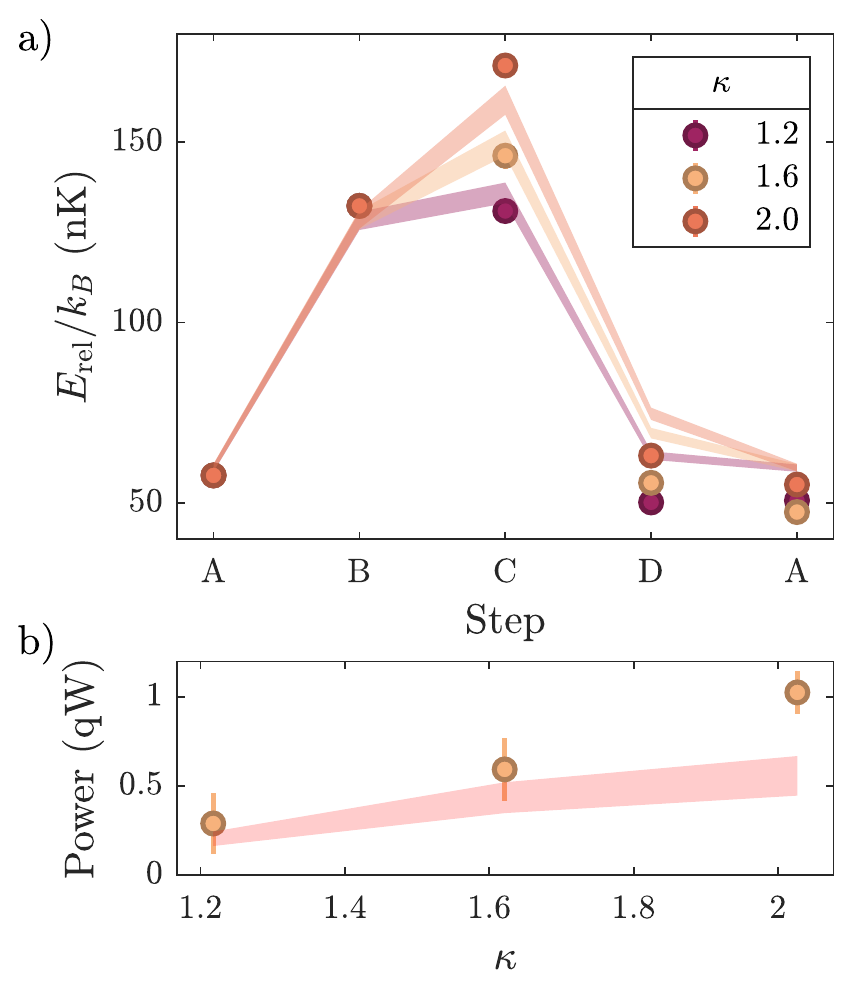}
\caption{Varying interaction strength ratio. \textbf{(a):} Measured release energy evolution over one engine cycle for varying interaction strength ratio $\kappa=a_s^C/a_s^A$ at a fixed compression ratio $\nu=1.94$. \textbf{(b):} Power output as a function of $\kappa$. Shaded regions in both panels are theoretical predictions from Eq.~\ref{eqn:erel_analytical} for the measured range of atom numbers.}
\label{fig5}
\end{figure}

Fig.~\ref{fig2}a shows close agreement between numerically calculated and experimentally measured release energies. This correspondence provides additional retroactive justification for the isentropic assumption, and also demonstrates that approximate analytic expressions describing only the condensate with $N_c$ particles provide relatively accurate estimates of the energy. The analytic formulas for release energy and total energy are~\cite{P_S_book}
\begin{equation}\label{eqn:erel_analytical}
\small
    \frac{\Erel}{N_c k_B T^0_c} = \frac{3 \zeta(4)}{2\zeta(3)} t^4 + \alpha \frac{1}{7}\left( (1-t^3)^{2/5}(2+\frac{17}{2}t^3)\right)
\end{equation}
and
\begin{equation}\label{eqn:etot_analytical}
\small
    \frac{\Etot}{N_c k_B T^0_c} = \frac{3 \zeta(4)}{\zeta(3)} t^4 + \alpha \frac{1}{7}\left( (1-t^3)^{2/5}(5+16t^3)\right),
\end{equation} with $\alpha=\mu_0/(k_B T_c^0)$, $\mu_0$ the zero-temperature chemical potential, $t=T/T_c^0$ the reduced temperature, $T_c^0$ the critical temperature of a harmonically confined Bose gas, $\zeta(x)$ the Riemann zeta function, and $k_B$ the Boltzmann constant. These results are accurate to within about 3\% of approximation-free numerical simulations at the measured condensate fraction.
The numerical results shown in Fig.~\ref{fig2}b do indicate that $\Ekin$ accounts for 10\% to 15\%  of the total energy in steps $A_1$ through $A_5$, violating to some extent the Thomas-Fermi approximation. Evaluating the ratios $P_\mathrm{tot}/P_\mathrm{rel}$ and $\eta_\mathrm{tot}/\eta_\mathrm{rel}$ using energies obtained from simulations shows that the former is 1\% to 3\% larger than predicted while the latter is 0.1\% to 0.6\% larger. 

A natural parameter to tune in order to maximize efficiency is the compression ratio $\nu = \omegabar^B/\omegabar^A$. Fig.~\ref{fig4}a shows measured release energy evolution over one cycle for different values of $\nu$. At higher compression ratios we observe distinctly higher release energies for steps \stb and \stc but no significant changes to the values at steps \sta and \std, in agreement with expectations from  Eqs.~\ref{eqn:erel_analytical} and \ref{eqn:etot_analytical}. Fig.~\ref{fig4}b demonstrates that increasing the compression ratio increases the efficiency $\eta$, which asymptotically approaches unity. In the Thomas-Fermi approximation this can be understood by analyzing the change in energy per particle $\Etot\propto a_s^{2/5}\omegabar^{\;6/5}$~\cite{pethick_smith_2008}. 
Defining $\kappa = a_s^C/a_s^A$ as the interaction ratio between steps \stc and \sta, the efficiency can be expressed as
\begin{equation} \label{eqn:effcomp}
    \eta=\frac{\kappa^{2/5}(\nu^{6/5}-1)-(\nu^{6/5}-1)}{\nu^{6/5}(\kappa^{2/5}-1)}=1-\nu^{-6/5}.
\end{equation}
As $\nu \rightarrow \infty, \eta \rightarrow 1$, and $\eta$ is independent of the interaction strength. This can be compared in loose analogy to the Otto cycle efficiency  $\eta_{\mathrm{Otto}}=1-\nu^{1-\gamma}$ with $\nu$ the compression ratio and $\gamma$ the specific heat ratio. 

Similarly, we can isolate the effects of interaction strength ratio $\kappa$ by holding the compression ratio constant. Following the same procedure used to derive Eq.~\ref{eqn:effcomp}, we find $P \propto (\kappa^{2/5}-1)(\nu^{6/5}-1)$: the power is determined solely by the interaction ratio $\kappa$ for a fixed compression ratio $\nu$. Fig.~\ref{fig5}a shows release energy evolution over one cycle for various interaction strength ratios corresponding to step \stc interaction strengths of 120, 160 and 200$a_0$, at a constant compression ratio $\nu=1.94$ and a particle number approximately 60\% larger than in Fig.~\ref{fig2}. Fig.~\ref{fig5}b shows that the output power indeed increases with $\kappa$, with a departure from theoretical predictions at larger values of $\kappa$ a possible hint of beyond-mean-field behavior. These results emphasize the importance of interaction effects in the engine: Feshbach tuning is the key parameter controlling energy transfer between magnetic and optical fields. This power enhancement is completely decoupled from the boost to efficiency achieved through stronger compression, and from the power enhancement due to decreased cycle time.  

In conclusion, we have realized an isentropic thermodynamic engine with a quantum degenerate working fluid and demonstrated that it outperforms a classical counterpart.
Experimental measurements of engine performance for various values of control parameters and degrees of adiabaticity are in good agreement with both low-temperature analytics and approximation-free numerical simulations. This work opens up a variety of interesting directions for future exploration. These include optimizing performance with shortcuts to adiabaticity~\cite{beau_scaling-up_2016,keller_feshbach_2020, li_efficient_2018,schaff_shortcut_2011}, realizing a quantum Otto refrigerator~\cite{hartmann_multispincounter_2020,niedenzu_quantized_2019,abah_stafridge_2020,jiao_ottofridge_2021,karimi_ottofridge_2016}, applying similar techniques to quantum heat engines involving trapped reservoirs of hot and cold atoms, investigating the role of criticality~\cite{chen_interaction-driven_2019}, and experimentally exploring the effects of entanglement on quantum thermodynamic engines~\cite{he_entangled_2012,zhang_four-level_2007,zhao_entangled_2017}.
\begin{acknowledgments}
We thank Kris Delaney and Ethan McGarrigle for theoretical contributions. D.W.\ acknowledges support from the National Science Foundation (2110584), the Air Force Office of Scientific Research (FA9550-20-1-0240), the Army Research Office (W911NF-20-1-0294),  and the Eddleman Center for Quantum Innovation, and from the NSF QLCI program through grant number OMA-2016245. G.F. acknowledges support from NSF DMR-2104255 for the theoretical method development.  R.S.\ and E.N.-M.\ acknowledge support from the UCSB NSF Quantum Foundry through the Q-AMASEi program (Grant No.\ DMR-1906325). Use was made of computational facilities purchased with funds from the National Science Foundation (CNS-1725797) and administered by the Center for Scientific Computing (CSC). The CSC is supported by the California NanoSystems Institute and the Materials Research Science and Engineering Center (MRSEC; NSF DMR 1720256) at UC Santa Barbara.
\end{acknowledgments}

\appendix
\section{Numerical Simulations}
To compose the full cycle in simulations, we must fix particle number, $N$, cell volume, $V$, and total entropy, $S$, around the cycle. All experimental observables are calculated by averaging field operators as described in the main text.
Operators for $N$, internal energy~\cite{G_K_book} and Helmholtz free energy~\cite{fredrickson_direct_2022} have been derived previously. Release energy is calculated using the operator for internal energy derived in~\cite{G_K_book}, excluding the contribution from the trap. Entropy $S$ is calculated from the Helmholtz free energy and internal energy. Note that this means $N$ is an output of the field theory, not a degree of freedom to be sampled, so the algorithmic cost is virtually independent of $N$. 

To perform the ensemble averaging, we allow the complex fields $\phi$ and $\phistar$ to independently evolve in a fictitous complex Langevin (CL) dynamics scheme according to a set of coupled stochastic partial differential equations that generates a Markov chain of system configurations~\cite{parisi, klauder}. Random noise correlations are chosen according to a fluctuation dissipation theorem~\cite{McQuarrie_book,  van_Kampen_book} that ensures that averages over CL time are equivalent to unbiased thermodynamic ensemble averages~\cite{gausterer_mechanism_1993, lee_convergence_nodate}, provided the CL dynamics have reached steady state prior to sampling. Although the operators may be complex, time or ensemble-average operators for physical observables are real.

We evolve the CL dynamics equations using the pseudospectral method detailed in~\cite{delaney_numerical_2020}, which decouples $\phi$ and $\phistar$ to linear order for numerical stability, and gives near-linear scaling with real space and $\tau$ resolution. We converge spatial resolution and imaginary time resolution until finite size effects in $\Erel$ and $S$ are no longer significant. For the simulations reported here, we use up to $160 \times 160 \times 128$ plane waves and 64 points in the $\tau$ direction. On an NVIDIA A100, the average simulation in continuous 3D space of approximately half a million particles at 170 nK converges to a time-independent solution in 2.5 hours, and by 24 hours statistical errors of the mean are less than 0.05\% of the mean. The longest simulations reported in this study had a duration of approximately 49 hours. 
 
We report only simulations computed with the average number of particles over the entire cycle. Initially, we performed two sets of simulations, one at $N=5 \times 10^5$ and one at $N=6 \times 10^5$, to account for experimental error in measured particle number of the data in Fig.~\ref{fig2}a. However, the range of simulation results was smaller than the line width in Fig.~\ref{fig2}a. Relative uncertainty in experimental $a_s$ and $\omega_i$ is smaller than the relative uncertainty in $N$, so we expect our results to be accurate for the reported experimental conditions. Cell volume $V$ is fixed such that the density distribution is well-contained within the simulation cell and finite size effects are no longer observed in the calculated release energy and entropy. For the largest sample, we used a simulation box of $71 \times 71 \times 57$ $\mu$m. $S$ is constrained by first computing the entropy at step \sta on the cycle using the experimental $T_A$ as an input parameter, then adjusting $T$ at all other points to maintain constant $S$. Using this procedure, $S$ remains within 2.5\% of its initial value in all cycles.

\section{Connection to the Otto Cycle}
\begin{figure}[hb!]
    \renewcommand{\figurename}{Fig.}
    \centering
    \includegraphics[scale = 1]{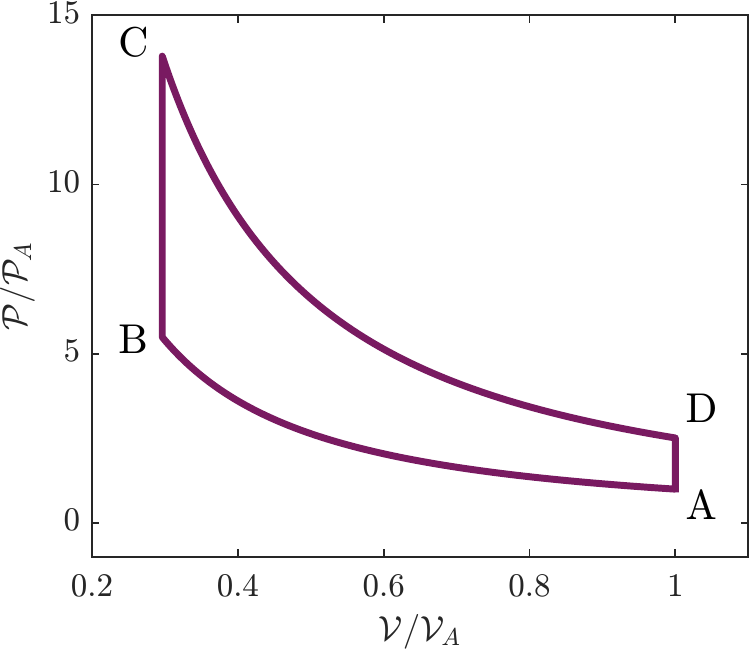}
    \caption{$\mathcal{PV}$ diagram for the thermodynamic engine. $\mathcal{V}_A$ and $\mathcal{P}_A$ are the harmonic volume and pressure evaluated at step \sta of the engine cycle. Here $\kappa = 10$ and $\nu = 1.5$. }
    \label{fig6}
\end{figure}

Here we draw an analogy between this isentropic thermodynamic cycle and the classical Otto cycle. First, following~\cite{romero-rochin_EOS_2005}, we define a ``harmonic volume'' $\mathcal{V} = (\hbar \bar \omega)^{-3}$ and write the total energy as
\begin{align}
 E = \frac{5}{7} \frac{15^{2/5}}{2} m^{1/5} N \left( \frac{N a_s}{\hbar \mathcal{V}} \right) ^{2/5}.
\end{align}

The ``harmonic'' pressure can then be derived using the fact that it is conjugate to volume:
\begin{align}
\mathcal{P} = -\frac{\partial E}{\partial \mathcal{V}}\bigg \vert_{N} = \frac{15^{2/5}}{7} m^{1/5} N \left( \frac{N a_s}{\hbar} \right) ^{2/5} \mathcal{V} ^{-7/5},
\end{align}

or equivalently by substituting the Thomas-Fermi density into the integral of the harmonic pressure given in~\cite{romero-rochin_EOS_2005}. The harmonic volume and pressure are not merely formal analogies; the harmonic volume can be associated with the physical volume that the particles occupy and the harmonic pressure can be associated with the mechanical equilibrium of the system. 

The total energy can then be rewritten as 
\begin{align}
E = \frac{5}{2} \frac{15^{2/5}}{7} m^{1/5} N \left( \frac{N a_s}{\hbar \mathcal{V}} \right) ^{2/5} = \frac{5}{2} \mathcal{PV},
\end{align}
and by using the definition of the Thomas-Fermi energy, we can recover an analogy to the ideal gas law:
\begin{align}
\mathcal{PV} = \frac{2}{7} N \mu. \label{eq:EOS}
\end{align}

It is important to note that while $\mu$ plays the role of an ``effective temperature'' it is unrelated to a thermal equilibrium. In our cycle, strokes of constant $\mu$ are analogous to isothermal strokes in the classical cycle. 

We now have all of the pieces to establish a connection with the Otto cycle. The first stage is an adiabatic compression $\bar \omega_A \rightarrow \bar \omega_B = \nu \bar \omega_A$ with compression ratio $\nu$. This traces an adiabat in the $\mathcal{PV}$-space.
Using  Eq.~\ref{eq:EOS}, the adiabat is defined by
\begin{align}
 \mathcal{V} \mu^{5/2} = \text{constant}, \quad \text{or}\quad \mathcal{V}^{7/5} \mathcal{P} = \text{constant}.
\end{align}
The heating stroke in the classical Otto cycle is replaced by an interaction strength stroke, which keeps the harmonic volume unchanged but changes the chemical potential and the harmonic pressure, thus mimicking an ``isochoric" process. We note that this is not an actual transfer of heat, as the thermodynamic entropy is constant. The final two strokes follow the same arguments presented above. A quantitative $\mathcal{PV}$ diagram of this thermodynamic cycle is shown in Fig.~\ref{fig6}.

This mathematical analogy enables an alternative derivation of the efficiency of the thermodynamic engine, allowing  us to use the Otto cycle efficiency directly with the adiabatic exponent $\gamma = 7/5$: 
\begin{align}
\eta = 1- \left( \frac{\mathcal{V}_B}{\mathcal{V}_A} \right)^{\gamma - 1} = 1- \left( \frac{\mathcal{V}_B}{\mathcal{V}_A} \right)^{2/5} \nonumber\\
= 1- \left( \frac{\bar\omega_A}{\bar \omega_B} \right)^{6/5} =1-\nu^{-6/5}
\end{align}
This is the same expression as Eq.~\ref{eqn:effcomp} in the main text.

\end{document}